\documentclass[11pt]{article}
 \pdfoutput=1 

\usepackage[english]{babel}
\usepackage[utf8x]{inputenc}
\usepackage[T1]{fontenc}

\usepackage[a4paper,top=3cm,bottom=2cm,left=3cm,right=3cm,marginparwidth=1.75cm]{geometry}

\usepackage{amsmath}
\usepackage{float}
\usepackage{graphicx}
\usepackage[colorinlistoftodos]{todonotes}
\usepackage[colorlinks=true, allcolors=blue]{hyperref}
\usepackage{authblk}
\usepackage{natbib}

\graphicspath{{figures/}}

\begin{document}

\title{Super Cyclone Amphan: A Dynamical Case Study}

\author[1]{Sridhar Balasubramanian\thanks{sridharb@iitb.ac.in}}
\author[2]{Vamsi K Chalamalla\thanks{vchalama@am.iitd.ac.in}}
\affil[1]{Department of Mechanical Engineering and IDP in Climate Studies, Indian Institute of Technology Bombay, Powai, India}
\affil[2]{Department of Applied Mechanics, Indian Institute of Technology Delhi, New Delhi, India}

\date{\today}  %\today is replaced with the current date

\maketitle

\begin{abstract}
Cyclone Amphan, a super cyclone in the Bay of Bengal after 21 years, intensified from a cyclonic storm (CAT 1) to a super cyclone (CAT 5) in less than 36 hours. It went on to make landfall over West Bengal as a Very Severe Cyclonic Storm (VSCS) with winds close to 155 kmph. Here, we analyze the dynamics that led to its rapid intensification, given that the system struggled to develop initially despite the presence of ripe conditions like high Sea Surface Temperature (SST) in the Bay. Our analysis clearly reveals that a Convectively Coupled Kelvin Wave (CCKW) from upper troposphere might have initiated strong instabilities in the tropopause, which then propagated vertically downward and interacted with surface disturbances to promote convective coupling with the Madden Julian Oscillations (MJO). Such convective coupling resulted in a burst of westerly winds along with enhanced vertical mixing and moisture convergence, which eventually led to the formation and intensification of super cyclone, Amphan.
\end{abstract}

%\copyrightstatement{TEXT}

\section{Introduction}
Atmosphere and ocean are replete with inertial waves and intra-seasonal oscillations, which play a very important role in transferring energy and momentum. The Madden Julian Oscillation (MJO) is the largest intra-seasonal variability, and air-sea interaction is one of the mechanisms contributing to MJO development and its propagation \citep{Flatau:1997,Flatau:2003,Fu:2004}. During the active phase of MJO, increased convective activity is observed leading to high cloudiness. The inertial waves are synoptic scale disturbances that propagate eastward or westward. The effects of MJO are further amplified by the presence of these inertial waves in the tropics. The most common waves that are seen near the Equator are the Kelvin wave and Rossby wave. In particular, Convectively coupled Kelvin waves (CCKWs) are an integral part of the equatorial dynamics in the atmosphere. The CCKWs are eastward moving waves on time scales ranging between 10-20 days and aid in the convective activity \cite{Kiladis:2009}. The CCKWs, together with other equatorial eastward and westward disturbances, connect the large-scale equatorial modes, such as Inter-tropical Convergence Zone (ITCZ) and MJO. The importance of CCKW was first recognized by \cite{Nakazawa:1988}, who noticed the eastward moving cloud ``superclusters'' embedded in the MJO envelope. CCKWs are separated from the MJO by their phase speed \citep{Straub:2002,Wheeler:1999}. While the phase speed of the MJO observed in the Indian Ocean is about 4–5 ms$^{−1}$, the phase speeds of convectively coupled Kelvin waves is $\approx$ 14 ms$^{−1}$ for suppressed MJO condition and $\approx$ 11 ms$^{−1}$ during the convective phase of the MJO \cite{Roundy:2008}. A recent field experiment (ASIRI-RAWI) in the Indian Ocean observed quasi-biweekly westerly wind bursts, which were linked to the upper level Kelvin waves that breakdown in the upper troposphere due to internal shear instabilities. These instabilities were found to interact with the surface disturbances to promote convective coupling with MJO \cite{Conry:2016}. \\

\cite{Frank_Roundy:2006} examined the role of tropical waves on tropical cyclogenesis and concluded that 
Kelvin waves has no significant role in the formation of tropical cyclones. \cite{Bessafi_Wheeler:2006} analyzed the modulation of tropical cyclones by various large-scale waves. They found a significant modulation of tropical cyclones by the MJO and Equatorial Rossby waves, and a small modulation by Kelvin waves. Another recent study by \cite{Schreck_Molinari:2011} examined the conditions leading to the formation of Typhoons Rammasun and Chataan. They observed an increase of the cyclonic potential vorticity with each passage of Kelvin wave. Overall, they found that both MJO and Kelvin waves played a role in the formation of these Typhoons. \cite{Mandke_Sahai:2016} investigated the role of equatorial waves in the formation of two tropical cyclones in the Indian Ocean. They found that convectively coupled Kelvin wave (CCKW) and convectively coupled equatorial Rossby wave (CCER) created favourable background conditions such as low-level westerlies and convection that led to the formation of twin cyclones in the Indian Ocean region. In view of these contradicting findings on the role of CCKWs on the formation or modulation of cyclones, it is important to further explore the role of Kelvin waves in the formation of tropical cyclones. \\

In this paper, we probe into the dynamics of Cyclone Amphan primarily looking at the interaction between CCKW, MJO, and the atmospheric stability/instability. To our knowledge, the effect of CCKWs on the intensification of tropical cyclones has not been looked into, which is the novelty that our study brings by taking Cyclone Amphan as a case study. The hypothesis we will be testing is the role of waves on the genesis and rapid intensification of cyclones.

\section{Formulation}

Convectively coupled Kelvin waves (CCKWs) are a component of atmospheric system that propagate eastward close to the equator. They are an important constituent of the convective envelope of MJO. It has been postulated that the ocean-atmosphere interactions within CCKWs may be important for MJO development, which could strongly impact the tropical climate, in general. In a field study named DYNAMO \citep{Matthews:2014}, CCKW events are strongly controlled by the MJO, with twice as many CCKWs observed during the convectively active phase of the MJO compared to the suppressed phase. Ocean and atmosphere coupling was observed during the passage of a CCKW, which lasts $\approx$ 4-5 days at any given longitude. Due the the passage of CCKWs, the westerly wind speed and latent heat flux are enhanced, leading to an enhanced convective phase of MJO.

\subsection {Governing equations and Dispersion Relation}
The solution for equatorial Kelvin waves can be derived from the linearized Navier Stokes equations using equatorially centered $\beta$ plane approximation as described below. Coriolis parameter $f$ is approximated as $f=\beta y$, where $y$ is the meridional or north-south coordinate.

\begin{align}
    \centering
	%\label{E01_first}
    & \frac{\partial{u}}{\partial{t}}  - \beta vy = -\frac{1}{\rho_0} \frac{\partial{p'}}{\partial{x}},
    \\
	& \frac{\partial{v}}{\partial{t}}  + \beta u y = -\frac{1}{\rho_0} \frac{\partial{p'}}{\partial{y}},
	\\
    & \frac{\partial{\rho'}}{\partial{t}} + w \frac{d \rho_0}{dz} = -\rho_0 (\nabla \cdot u),
    \\
    & \frac{\partial{p'}}{\partial{t}} -\rho_0 gw = c_s^2(\frac{\partial{\rho'}}{\partial{t}} + w \frac{d \rho_0}{dz}).
\end{align}

Equatorial Kelvin waves are eastward propagating waves with only zonal component of velocity. Assuming wave-like solutions for all quantities i.e $e^{i(kx+\omega t)}$, dispersion relation for equatorially trapped waves can be obtained. Kelvin waves are non-dispersive with zonal velocity having Gaussian meridional structure. The wave speed for equatorial Kelvin wave is given by $\sqrt{gH_e}$, where $g$ is the acceleration due to gravity and $H_e$ is the equivalent depth which depends on the background conditions under which the wave exists. Convectively Coupled Kelvin Waves (CCKWs) have all the basic features of an equatorial Kelvin wave in addition to the tight coupling with convection. Typically CCKWs were observed to have a wave speeds ranging from $10$ m/s to $30$ m/s resulting in an equivalent depth $H_e$ ranging between 10--100 m.

% \vspace{-1em}
%\begin{table}[H]
%  \centering
%  \renewcommand{\arraystretch}{1.25}
%  \setlength{\tabcolsep}{4pt}
%  \begin{tabular}{| c || c | c | c | c | c | c | c | c | c |p{1cm} |}
%   \hline
%    \bf{Case} & $\bf{L_x (m)}$ & $\bf{L_y (m)}$  & $\bf{L_z (m)}$ & $\bf{N_x}$ & $\bf{N_y}$  & $\bf{N_z}$ & $\bf{dx (m)}$ & $\bf{dy (m)}$  & $\bf{dz (m)}$ \\ \hline \hline
%    1 & 4096 & 4096 & 128 & 512 & 512 & 64 & 8 & 8 & 2  \\ \hline
%  \end{tabular}
%  \caption{Grid parameters. $N_x,N_y,N_z$ in this table represent number of grid points on the coarse level. }
%  \label{tab:grid_parameters}
%\end{table}

 \bigskip 

\section{Result and Analysis}
We analyze the dynamics of super cyclone Amphan using the Sea Surface Temperature (SST), air temperature, wind speed, and velocity potential at 200mb level data sets. Additionally, we perform a comparative study of Amphan with other recent cyclones in this region, namely, cyclone Fani and cyclone Ockhi to reveal the role of CCKWs in Amphan's genesis and intensification.

\subsection{Super cylcone Amphan} 

Cyclone Amphan was the first super cyclonic storm to occur in the Bay of Bengal since the 1999 Odisha cyclone. It made landfall as a Very Severe Cyclonic Storm (VSCS) in West Bengal on May 20, 2020. What made Amphan so deadly was its rapid intensification from a Cyclonic Storm (CS) to a super cyclone in less than 36 hours. Below, we analyze the probable dynamical reasons for the rapid intensification of Amphan. \\ 

\begin{figure}
\includegraphics[width=11cm]{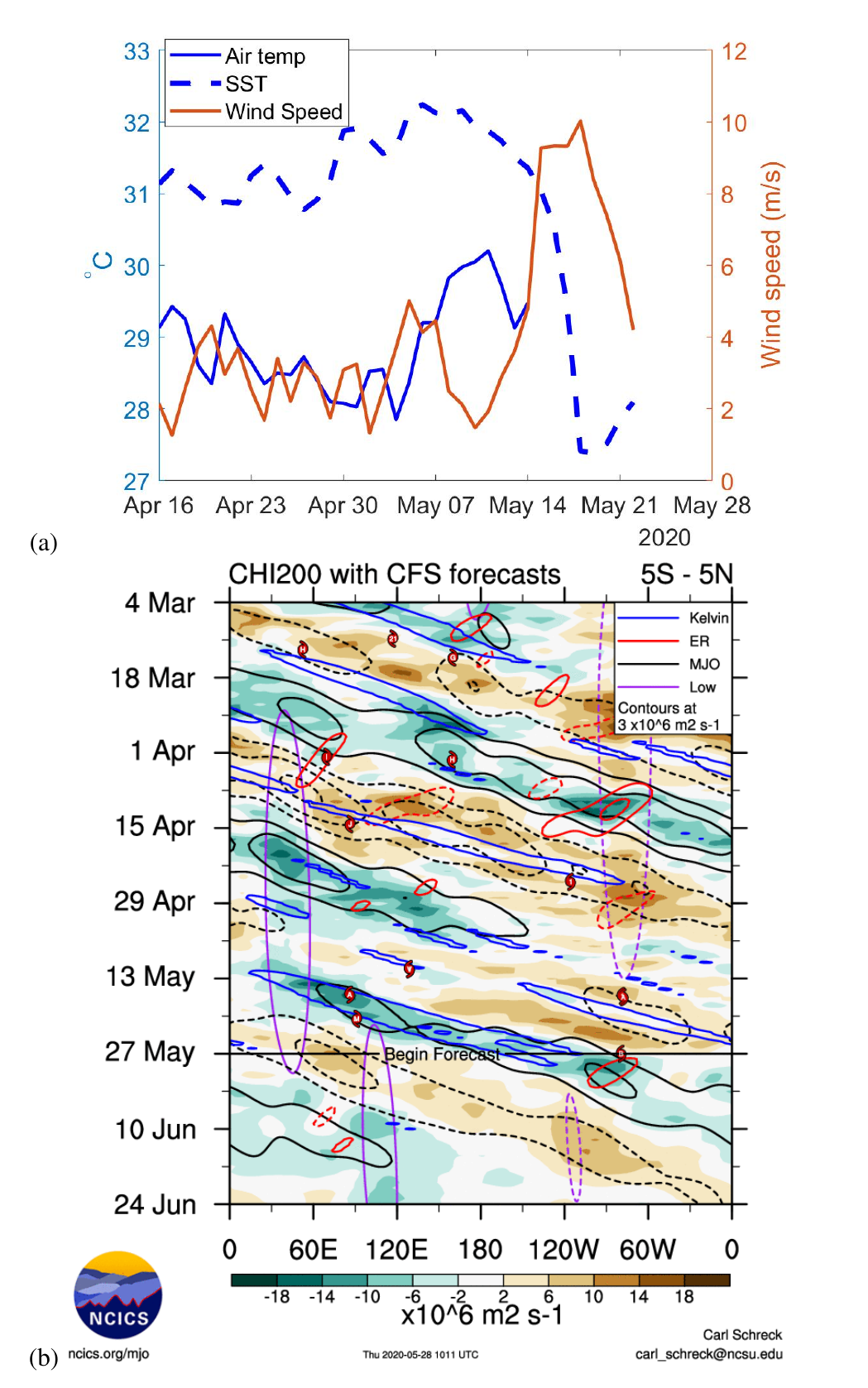} \\
\caption{(a) Daily averaged air temperature, sea surface temperature (SST), and wind speed near buoy location BD14 (7N/88E) one month prior to the origin of cyclone Amphan,(b) Hovmöller diagram of velocity potential at 200 hPa averaged between $5^\circ$ S and $5 ^\circ$ N.}
\label{fig1}
\end{figure}

\begin{figure}
\includegraphics[width=13cm]{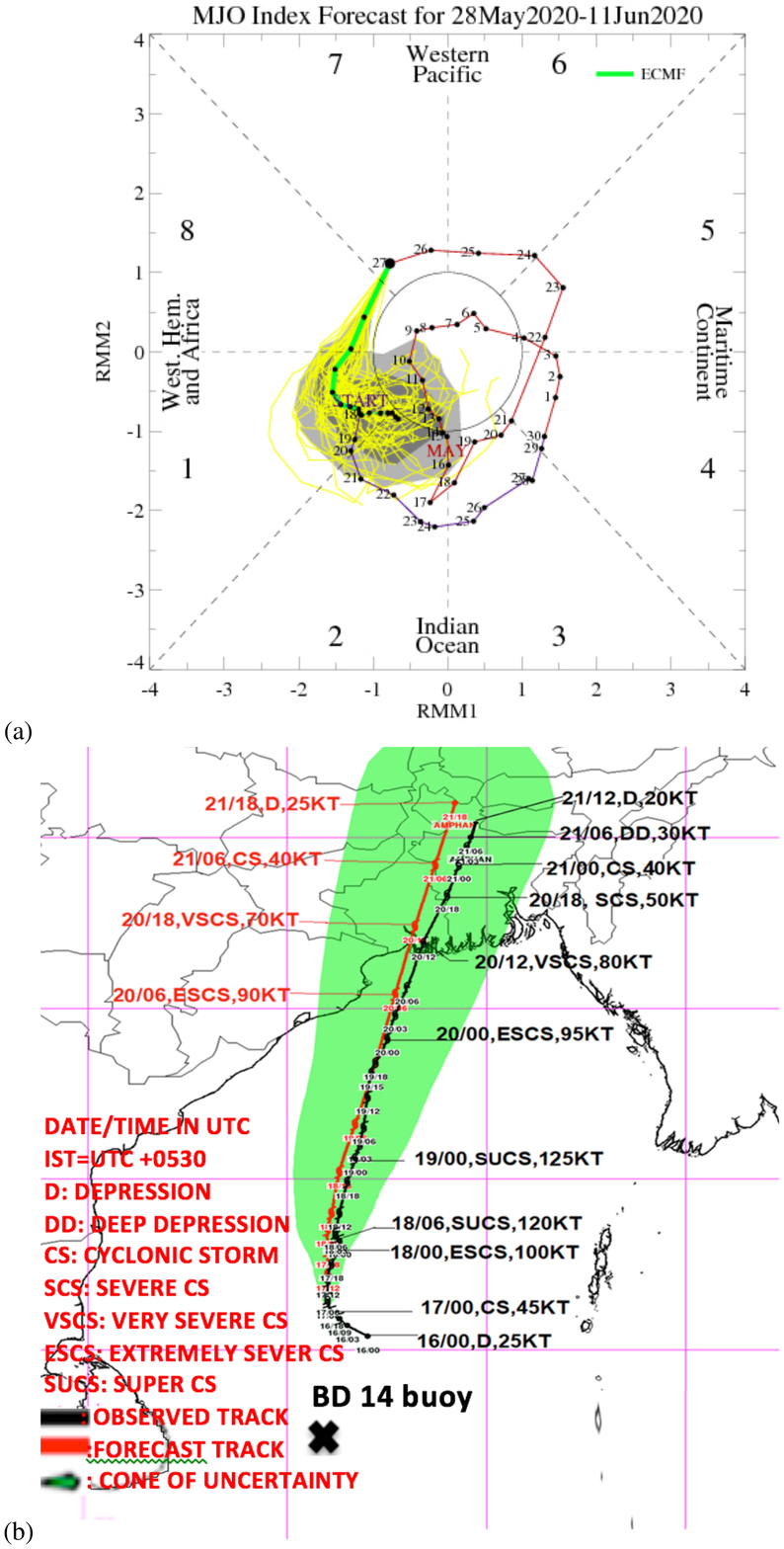} \\
\caption{(a) MJO phase plot, (b) path of cyclone Amphan.}
\label{fig2}
\end{figure}

The SST, air temperature at 2m level, and wind speed from buoy BD14 (7N/88E) are shown in Fig. \ref{fig1}a. It is observed that for the month of April and May, the SST maintains a very high value of $\approx$ 31--32 $^{\circ}$C. Note that, the drop in SST after May 16 at BD14 is due to the upwelling of cold water to the surface as a result of formation of cyclone Amphan. Looking at the wind speed in Fig. \ref{fig1}a, we see a spike in the surface level winds around May 14. The wind speeds during this spike is $\approx$ 10--11 ms$^{-1}$, which is the typical speed of a CCKW observed in the Indian Ocean. Thus, one could argue that the increase in the wind speed around May 14, 2020 is due to the CCKW that made its entry into the East Indian Ocean. Moreover, genesis point of cyclone Amphan was farther north of the buoy BD14 as shown in Fig. \ref{fig2}b. Finally, Amphan developed into a depression/cyclone on May 16, 2020 and it never crossed or came close to BD14 buoy. The fact that the winds are seen increasing at BD14 buoy from May 14 is a clear signature of CCKW.

Hovmöller diagram of velocity potential at 200 hPa averaged between $5^\circ$ S and $5 ^\circ$ N is shown in Fig. \ref{fig1}b (taken from https://ncics.org/portfolio/monitor/mjo/). Different contours seen in the plot correspond to MJO, equatorial Rossby waves, Kelvin waves, and other low frequency components. The algorithm developed by \cite{Wheeler_Weickmann:2001} for filtering outgoing longwave radiation (OLR) in real-time is used to identify the MJO and equatorial waves. This filtering technique involves FFT analysis of OLR anomaly data with zero-padding for the specific zonal wavenumbers and frequencies. Typically, filtering in the time period range of 2.5--20 days, with eastward wave numbers 1--14 is performed to identify CCKWs. This methodology successfully captures CCKWs in the real-time analysis.

The fact that the velocity potential at 200 hPa is negative indicates divergence in the upper level and convergence in low levels, a signature of deep convection event. Also, Kelvin waves are generated near Africa prior to the formation of Amphan (see blue contours in Fig. \ref{fig1}b), which then traveled to the East Indian Ocean. So, the signature of deep convection along with active Kelvin waves as shown in the Hovmöller diagram further confirm the role of CCKWs in the genesis and intensification of cyclone Amphan.

The MJO phase plot in Fig. \ref{fig2}a shows an increase in the amplitude between May 16 and 18 while straddling between Phases 2 and 3. This spike is due to the interaction of MJO with the strong CCKW present during this period, which allowed the MJO signal to amplify (Roundy, 2008). 

From this analysis we conclude that the CCKW was the primary reason for the genesis and rapid intensification of cyclone Amphan. The high SST only played a secondary role, since the SSTs were high for quite some time (since April 2020 as seen from Fig. \ref{fig1}a), but the system was struggling to intensify for many days up until the arrival of the CCKW, which gave the much needed momentum for the system to intensify. In order to further drive home the point regarding the role of CCKW during Amphan, we compare it with two other cyclones, namely, Fani and Ockhi.

\subsection{Cyclone Fani} 
\begin{figure}
\includegraphics[width=13cm]{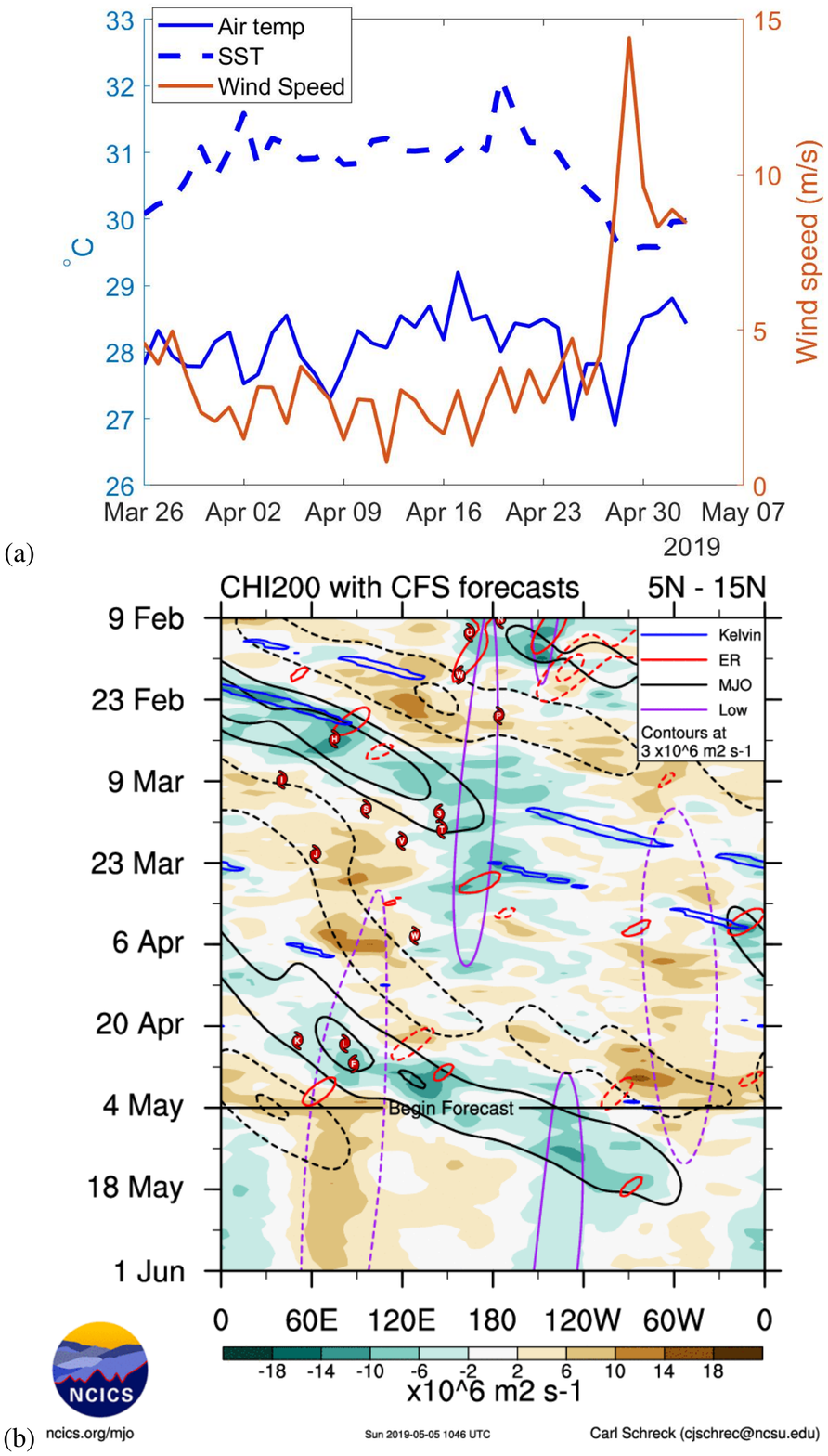} \\
\caption{(a)Daily averaged air temperature, sea surface temperature (SST), and wind speed near buoy location BD14 (7N/88E) one month prior to the origin of cyclone Fani ,(b) Hovmöller diagram of velocity potential at 200 hPa averaged between $5^\circ$ S and $5 ^\circ$ N.}
\label{fig3}
\end{figure}

\begin{figure}[t]
\includegraphics[trim={0 5cm 0 0},width=13cm]{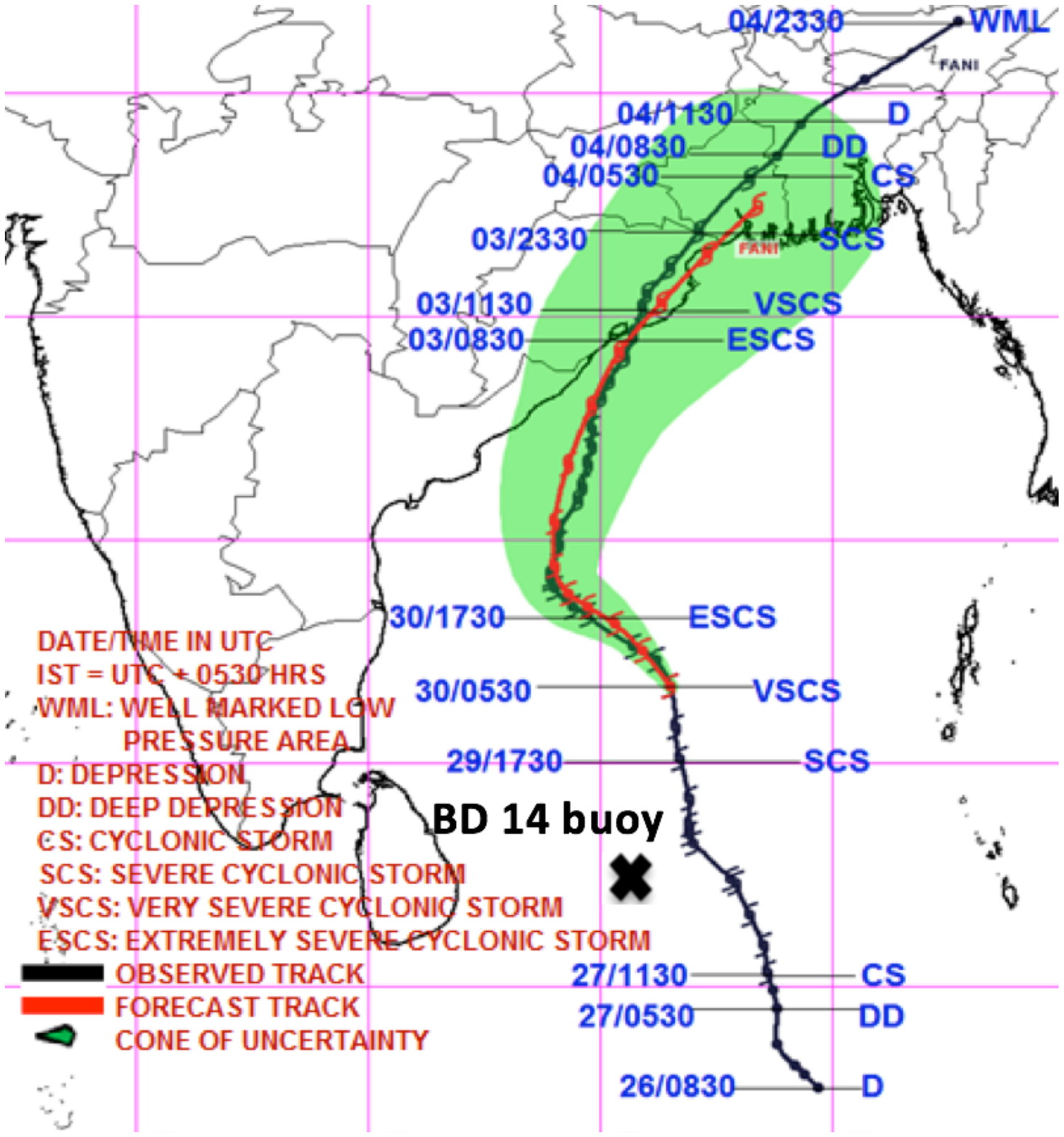} \\
\caption{Path of cyclone Fani.}
\label{fig4}
\end{figure}

Cyclone Fani was a Extremely Severe Cyclonic Storm (ESCS) to form in the Bay of Bengal, which went on to make landfall over Odisha on May 3, 2019. This made Fani the most intense storm to make landfall in Odisha since the 1999 Odisha cyclone. Despite being a strong system, Fani showed gradual intensification unlike Amphan. Moreover, it never became a super cyclonic storm, despite the fact that the Bay of Bengal basin had very high SST during both the cyclones, i.e. higher than normal SST (31-32 C) recorded by buoy BD14 during April 2019 and May 2020. This shows that the SST alone was not the driver for Super Cyclone Amphan and indeed it was the CCKW responsible for its rapid intensification to a Super Cyclone.

From the BD14 buoy data in Fig. \ref{fig3}a, it is observed that the wind speeds are much lower than the typical CCKW speeds of $\approx$ 11 ms$^{-1}$ in the Indian Ocean, before the formation of the cyclone. The wind speed is around 4 ms$^{-1}$, which is typical during an active MJO phase. There is an increase in the wind speed around April 28, which is attributed to the fact that Fani crossed this buoy on that day (see Fig. \ref{fig4}). Hence, there was no CCKW presence during the genesis of Fani. It was fully governed by the air-sea coupling or strong MJO. This is also confirmed from the velocity potential at 200 mb plot (see Fig. \ref{fig3}b). This comparison shows that the role of CCKW during Amphan cannot be ignored and this was the reason for Amphan's rapid intensification and not high SST values.

\subsection{Ockhi} 

In order to drive home the point of CCKW, we investigated another case study, cyclone Ochki. It was a Very Severe Cyclonic Storm (VSCS) that impacted Peninsular India. Ockhi garnered so much attention, because of its rapid intensification from a Depression to a Cyclonic Storm. 

\begin{figure}
\includegraphics[width=13cm]{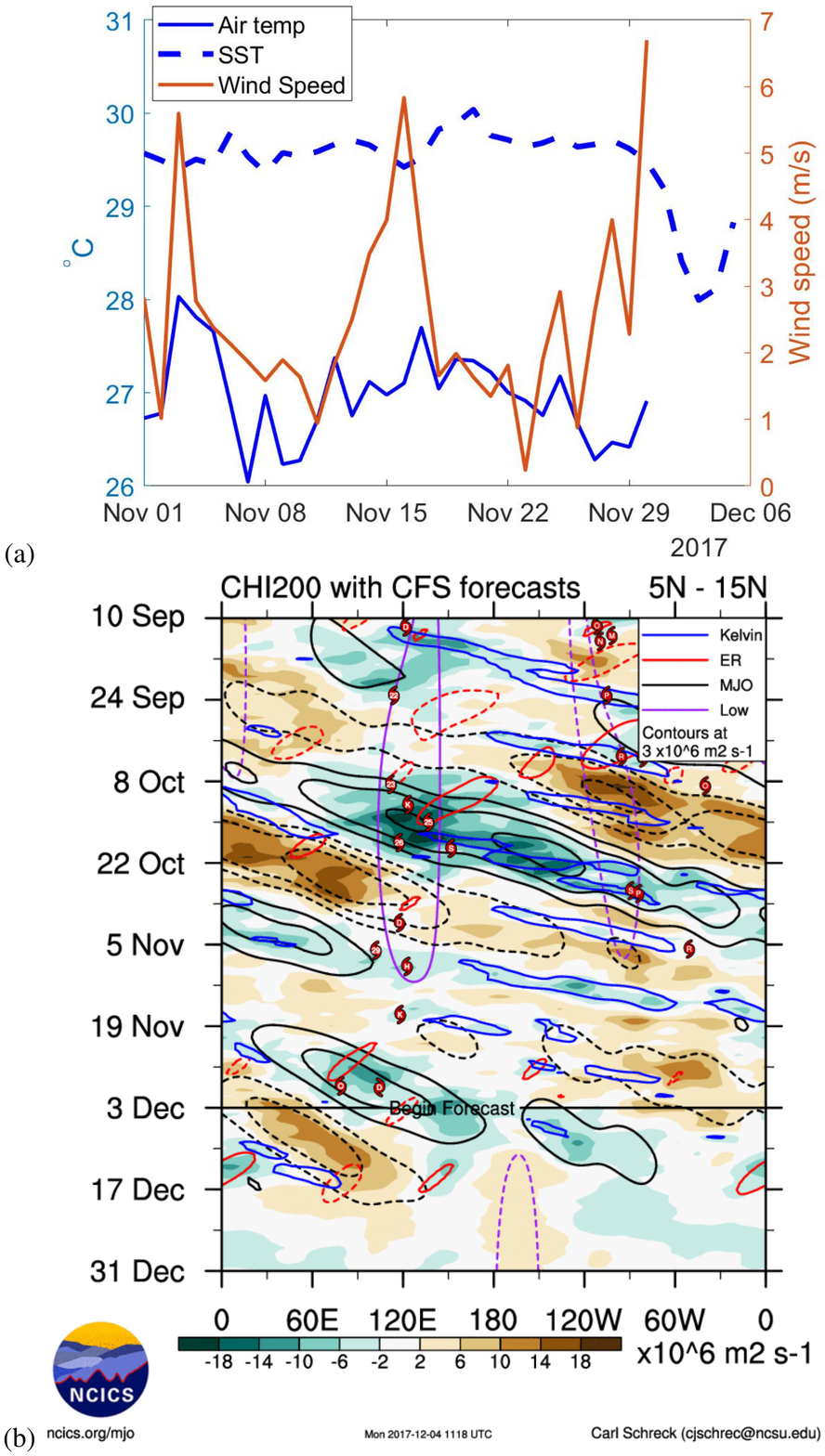} \\
\caption{(a)Daily averaged air temperature, sea surface temperature (SST), and wind speed near buoy location AD09 (8N/73E) one month prior to the origin of cyclone Ockhi,(b) Hovmöller diagram of velocity potential at 200 hPa averaged between $5^\circ$ S and $5 ^\circ$ N.}
\label{fig5}
\end{figure}

\begin{figure}
\centering
\includegraphics[trim={0 8cm 0 7cm},width=13cm]{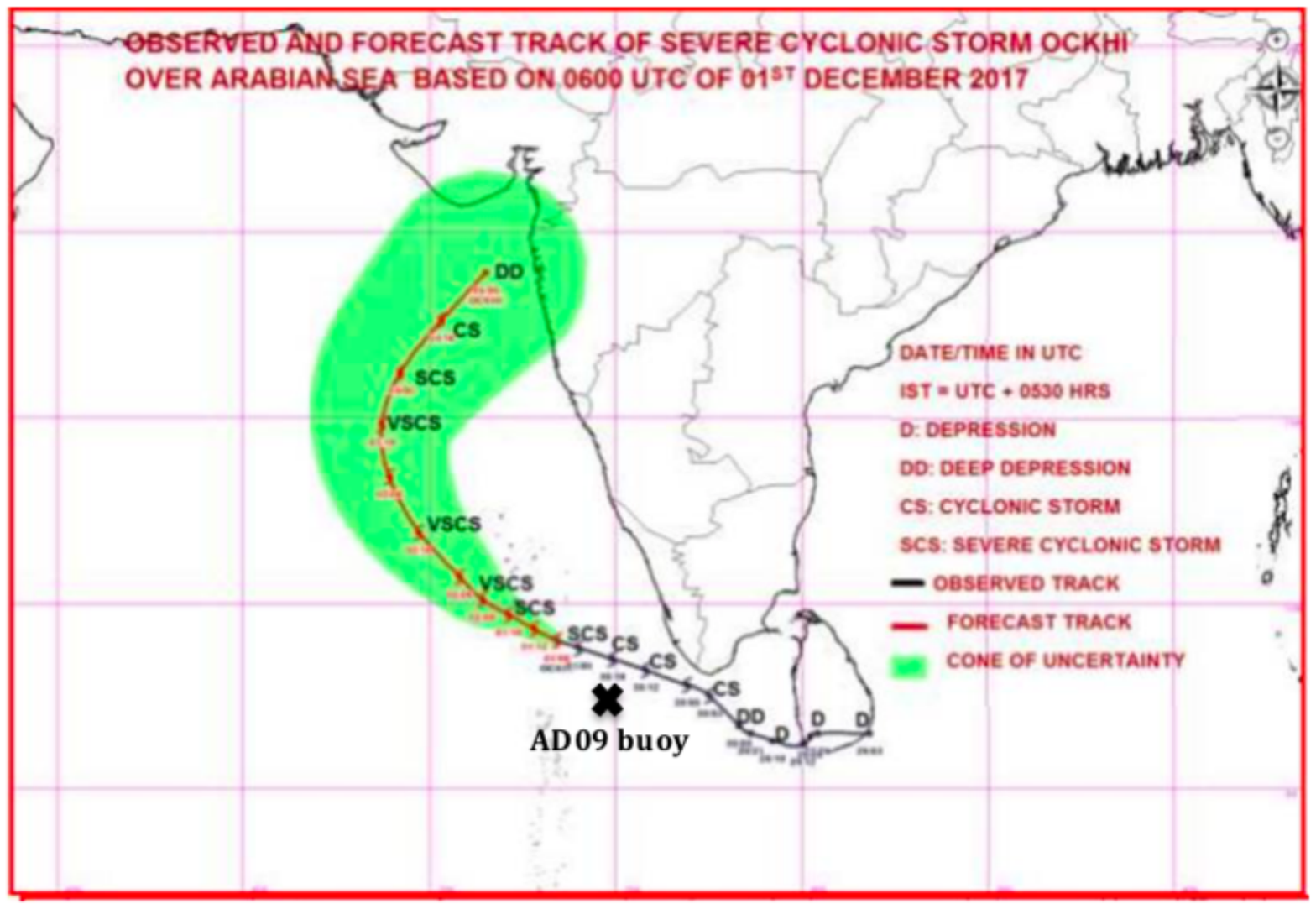} \caption{Path of cyclone Ockhi.}
\label{fig6}
\end{figure}

Looking at the buoy data AD09 in the Arabian Sea, we see that the SST is between 29-30$^{\circ}$C as shown in Fig. \ref{fig5}a. This is higher than the normal SST values observed in the Arabian Sea during the month of December. The high SST led to air-sea coupling and resulted in a strong MJO pulse to sustain in the Arabian Sea, which led to intensification of Ockhi (see Fig. \ref{fig5}b, which confirmed MJO pulse). If we look at the wind speed during Ockhi, it is much lower than the typical CCKW wind speed of $\approx$ 11 ms$^{-1}$ in the Indian Ocean, before the genesis of the cyclone. We see an increase around November 30, which is again due to the fact that Ockhi crossed AD09 buoy on that day (see Fig. \ref{fig6}). Hence, the role of CCKW can be ruled out during Ockhi.

To further understand the reasons behind intensification of cyclone Amphan, we plotted the equivalent potential temperature $\theta_e$ as a function of depth to quantify the extent of convectively unstable regions. The NCEP data was used for this purpose. The value of $\theta_e$ is calculated using the following equation:

\begin{equation}
    \theta_e\approx\theta \exp(1+\frac{Lw}{C_pT})
\end{equation}
where $\theta$ is the potential temperature and $L$ is the latent heat, $C_p$ is the specific heat, $w$ is the mixing ratio and $T$ is the air temperature. The primary importance of $\theta_e$ is in terms of stability of the atmosphere. If $\frac{d\theta_e}{dz}<0$, then the atmosphere is unstable, thereby allowing for high degree of mixing. When $\frac{d\theta_e}{dz}>0$, the atmosphere is stable inhibiting the mixing. \\

\begin{figure}
\centering
 {\small{(a)}}
\includegraphics[width=0.7\textwidth]{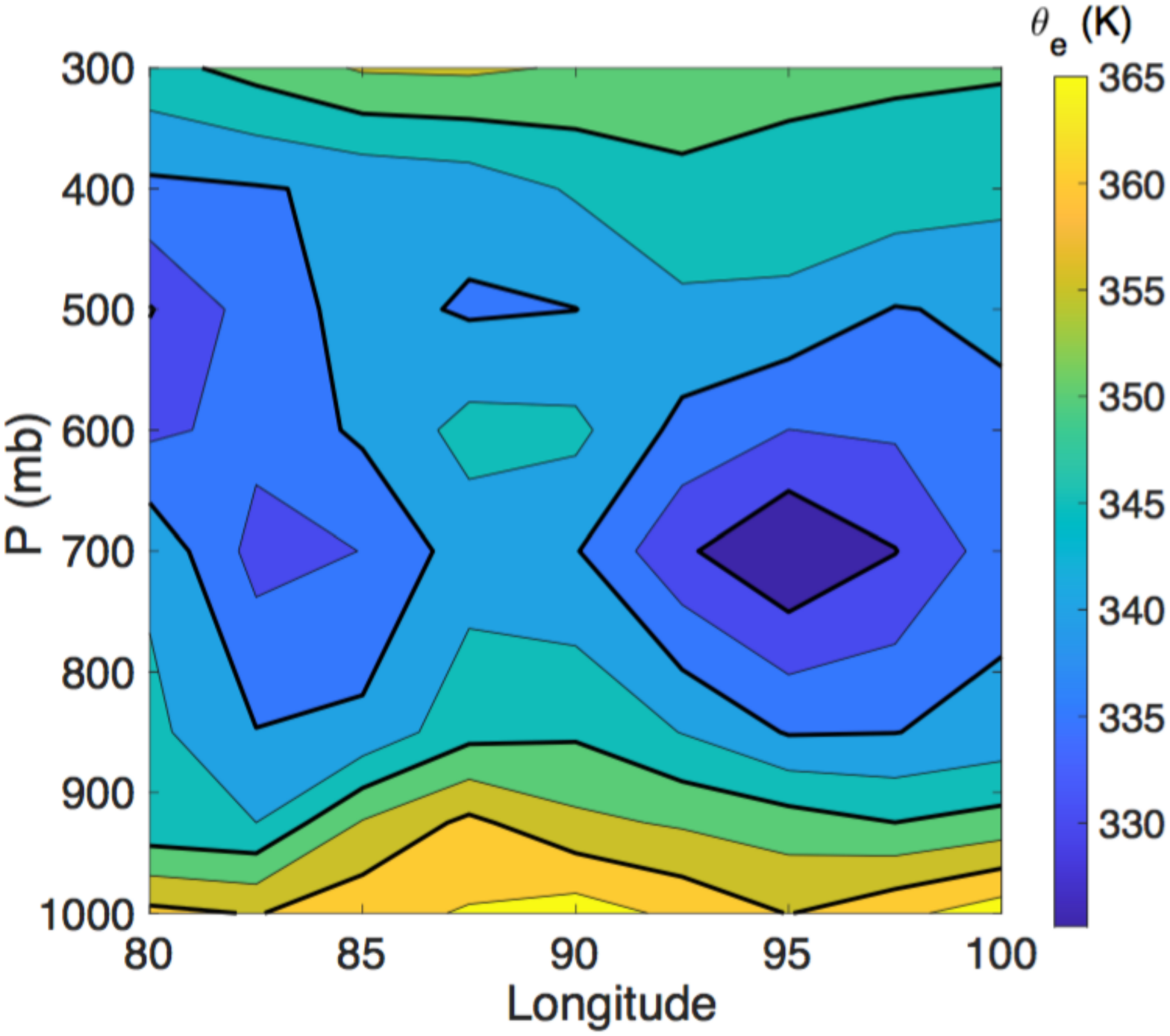} \\
{\small{(b)}}
\includegraphics[width=0.85\textwidth]{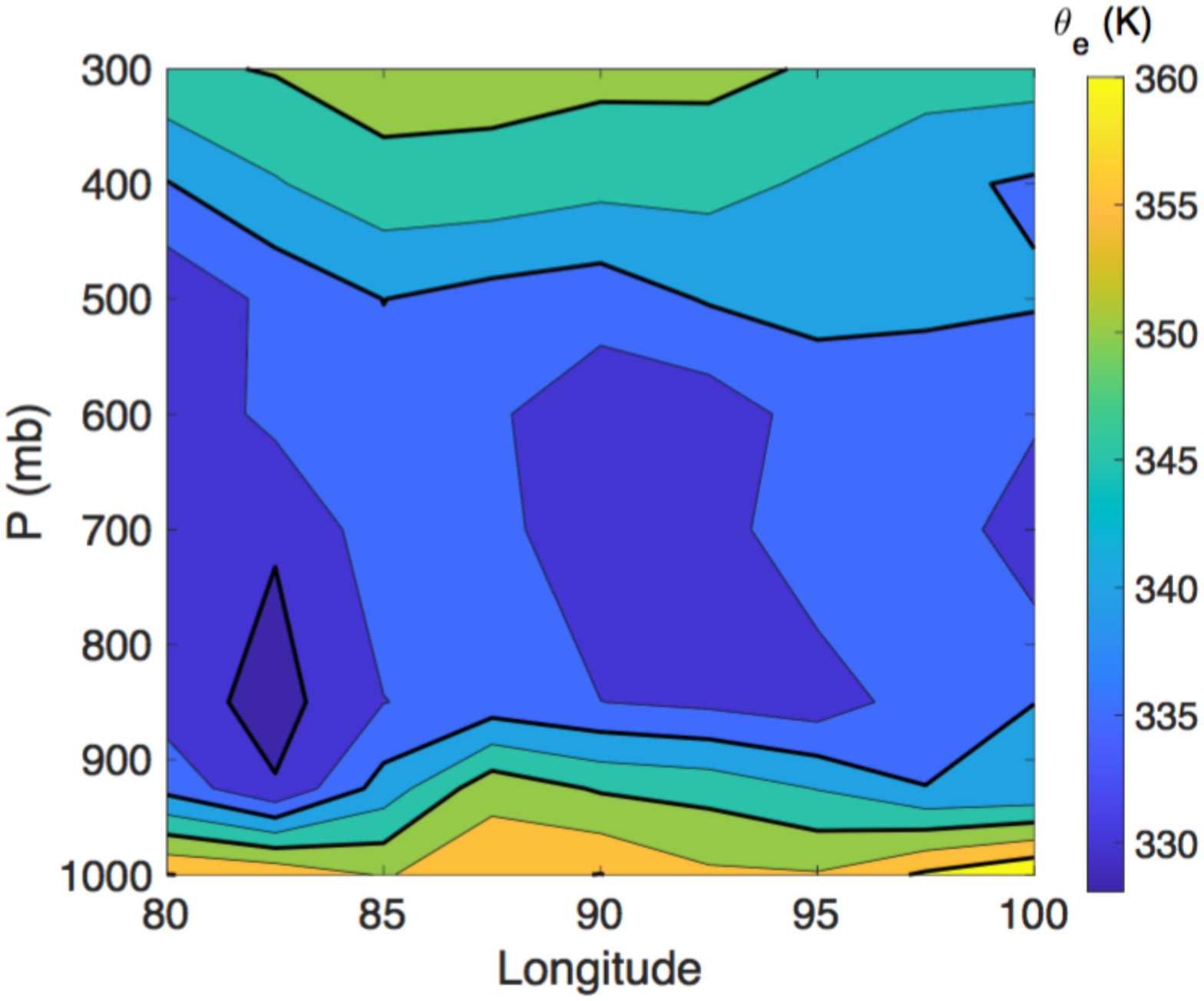} \\
\caption{Contour plots of equivalent potential temperature at $12.5 ^\circ$ latitude corresponding to (a) Cyclone Amphan on 18/05/2020, and (b) Cyclone Fani on 30/04/2019.}
\label{pot_temp_contours}
\end{figure}

\begin{figure}
\includegraphics[trim={0 11cm 0 10cm},clip,width=18cm]{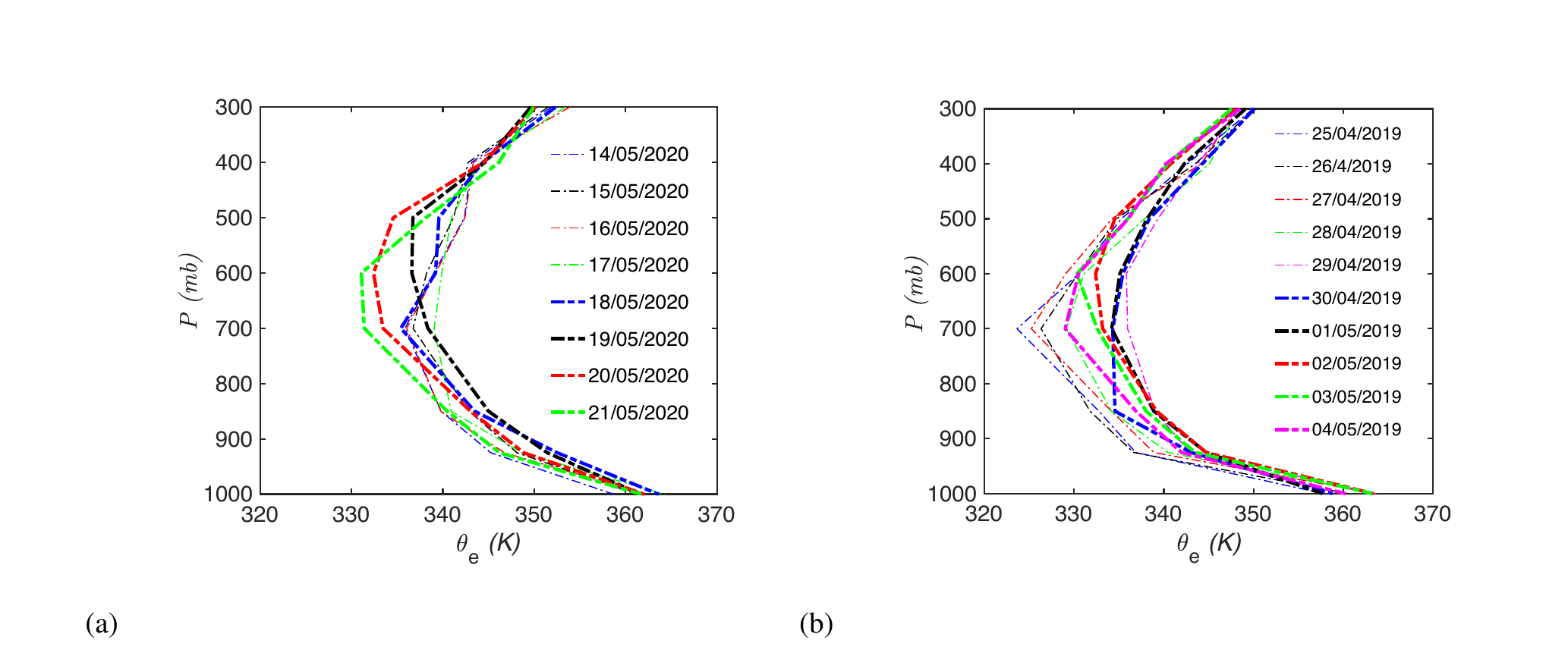} \caption{Longitudinally averaged vertical profiles of equivalent potential temperature at $12.5 ^\circ$ latitude corresponding to (a) Cyclone Amphan, (b) Cyclone Fani. } 
\label{pot_temp_lines}
\end{figure}

Figure \ref{pot_temp_lines} shows longitudinally averaged equivalent potential temperature at 12.5$^\circ$ latitude  for various days corresponding to each cyclone. Both the cyclones show a similar qualitative trend i.e, $\theta_e$ decreases from the surface followed by an increase. In the case of Amphan, minimum $\theta_e$ is observed at $\approx$ $500$ mb level, whereas in the case of Fani, minimum temperature is observed at $\approx$ $700$ mb. The region of unstable motion (which is a surrogate for mixing and entrainment) for the case of cyclone Amphan extends up to mid-troposphere. This clearly indicates that the region of instability extended up to large altitudes in the case of cyclone Amphan when compared to Fani. Also, the longitudinal variation of $\theta_e$ is small in both cases as seen from the contour plots of $\theta_e$ in Fig. \ref{pot_temp_contours}, suggesting that longitudinal average of $\theta_e$ is a good indicator of instability associated with each configuration. These figures collectively show that the extent of unstable region and convection is larger for cyclone Amphan when compared to Fani. The extended zone of instability could be attributed to intense convective activity and vertical moisture updraft for cyclone Amphan. It has been documented that the CCKW disturbances from upper atmosphere travel vertically downwards to cause intense instabilities in the lower and mid tropospheric levels \cite{Conry:2016}. The CCKW are also associated with a burst of westerly moisture flux. The presence of a strong CCKW from Fig. \ref{fig1} in conjunction with Fig. \ref{pot_temp_lines}a confirms that the CCKW had a role to play in enhancing the moisture convergence and vertical mixing, thereby possibly resulting in a rapid amplification of the cyclonic storm.

From the three case studies presented here, we clearly see that the CCKW played a vital role in the rapid intensification of Cyclone Amphan to a super cyclone. This further goes to show that the SST and ocean heat content are only some of the factors that play a role in cyclone intensification and probably should not be solely used to explain the rapid intensification of cyclones to super cyclones. The role of convective coupled waves, like CCKW, and the associated atmospheric instability should also be considered. Our dynamical case study shows the important role of Convectively Coupled Kelvin Waves (CCKW) on the genesis and intensification of tropical cyclones, which must be accounted for while predicting cyclones or for that matter any tropical system.
%\bigskip 

\section{Conclusion}
In this study, we analyzed the dynamics behind super cyclone Amphan, focusing on the possible reason for its rapid intensification to a super cyclone. We document that a strong Convectively Coupled Kelvin Wave (CCKW) interacted with the MJO, resulting in a momentary strong MJO in the East Indian Ocean basin. The presence of CCKW lead to a strong westerly moisture burst that acted as a constant source of moisture for this system. The diabatic heating as a result of the moisture flux resulting in an atmospheric instability extending up to mid troposphere, thereby enhancing the vertical mixing and updraft. We conclude that high SST was not the primary and sole reason for the genesis and intensification of Amphan, as has been documented in many media reports. The role of the strong CCKW could not be ignored and CCKW could be a potential reason for Amphan becoming a super cyclone in a record time of less than 36 hours. Our study emphasizes the role of inertial waves in tropical cyclones and other synoptic weather systems in the tropics.

\section{Acknowledgment}
We gratefully acknowledge the open source data and maps from NCICS, IMD, INCOIS, NCEP, and ECMWF, which were used in this study.

\bibliographystyle{plain}
\bibliography{main.bib}

\begin{thebibliography}{15}
\providecommand{\natexlab}[1]{#1}
\providecommand{\url}[1]{{\tt #1}}
\providecommand{\urlprefix}{URL }
\expandafter\ifx\csname urlstyle\endcsname\relax
  \providecommand{\doi}[1]{https://doi.org/\discretionary{}{}{}#1}\else
  \providecommand{\doi}{https://doi.org/\discretionary{}{}{}\begingroup
  \urlstyle{rm}\Url}\fi

\bibitem[{Bessafi and Wheeler(2006)}]{Bessafi_Wheeler:2006}
Bessafi, M. and Wheeler, M.: Modulation of south Indian Ocean tropical cyclones
  by the Madden-Julian Oscillation and convectively coupled equatorial waves,
  Mon. Wea. Rev., 134, 638--656, 2006.

\bibitem[{Conry et~al.(2016)Conry, Fernando, Leo, Blomquist, Amelie, Lalande,
  Creegan, Hocut, MacCall, Wang, Jinadasa, Wang, and Yeo}]{Conry:2016}
Conry, P., Fernando, H. J.~S., Leo, L., Blomquist, B., Amelie, V., Lalande, N.,
  Creegan, E., Hocut, C., MacCall, B., Wang, Y., Jinadasa, S. U.~P., Wang, C.,
  and Yeo, L.~K.: Observations of Equatorial Kelvin Waves and their Convective
  Coupling with the Atmosphere/Ocean Surface Layer, APS Division of Fluid
  Dynamics (Fall) 2016, 2016APS..DFD.A5004C, 2016.

\bibitem[{Flatau et~al.(1997)Flatau, Flatau, Phoebus, and Niller}]{Flatau:1997}
Flatau, M., Flatau, P.~J., Phoebus, P., and Niller, P.~P.: The feedback between
  equatorial convection and local radiative and evaporative processes: The
  implications for intraseasonal oscillations, J. Atmos. Sci., 54, 2373--2386,
  1997.

\bibitem[{Flatau et~al.(2003)Flatau, Flatau, Schmidt, and
  Kiladis}]{Flatau:2003}
Flatau, M.~K., Flatau, P.~J., Schmidt, J., and Kiladis, G.~N.: Delayed onset of
  the 2002 Indian monsoon, Geophys. Res. Lett., 30, 1768, 2003.

\bibitem[{Frank and Roundy(2006)}]{Frank_Roundy:2006}
Frank, N.~L. and Roundy, P.~E.: The role of tropical waves in tropical
  cyclogenesis, Mon. Wea. Rev., 134, 2397--2417, 2006.

\bibitem[{Fu and Wand(2004)}]{Fu:2004}
Fu, X.~H. and Wand, B.: Differences of boreal summer intraseasonal oscillations
  simulated in an atmosphere-ocean coupled model and an atmosphere-only
  model, J. Clim., 17, 1263--1271, 2004.

\bibitem[{Kiladis et~al.(2009)Kiladis, Wheeler, Haertel, Straub, and
  Roundy}]{Kiladis:2009}
Kiladis, G.~N., Wheeler, M.~C., Haertel, P.~T., Straub, K.~H., and Roundy,
  P.~E.: Convectively coupled equatorial waves, Rev. Geophys., 47, RG2003,
  2009.

\bibitem[{Mandke and Sahai(2016)}]{Mandke_Sahai:2016}
Mandke, S.~K. and Sahai, A.: Twin tropical cyclones in the Indian Ocean: the
  role of equatorial waves, Nat. Hazards, 84, 2211--2224, 2016.

\bibitem[{Matthews et~al.(2014)Matthews, Baranowski, Heywood, Flatau, and
  Sunke}]{Matthews:2014}
Matthews, A.~J., Baranowski, D.~B., Heywood, K.~J., Flatau, P.~J., and Sunke,
  S.: The surface diurnal warm layer in the Indian Ocean during CINDY/DYNAMO,
  J. Clim., 27, 9101--9122, 2014.

\bibitem[{Nakazawa(1988)}]{Nakazawa:1988}
Nakazawa, T.: Tropical super clusters within intraseasonal variations over the
  western Pacific, J. Meteorol. Soc. Jpn., 66, 823--839, 1988.

\bibitem[{Roundy(2008)}]{Roundy:2008}
Roundy, P.~E.: Analysis of convectively coupled Kelvin waves in the Indian
  ocean MJO, J. Atmos. Sci., 65, 1342--1359, 2008.

\bibitem[{Schreck and Molinari(2011)}]{Schreck_Molinari:2011}
Schreck, C. J.~I. and Molinari, J.: Tropical cyclogenesis associated with
  Kelvin Waves and the Madden-Julian Oscillation, Mon. Wea. Rev., 139,
  2723--2734, 2011.

\bibitem[{Straub and Kiladis(2002)}]{Straub:2002}
Straub, K.~H. and Kiladis, G.~N.: Observations of a convectively coupled Kelvin
  wave in the eastern Pacific ITCZ, J. Atmos. Sci., 59, 30--53, 2002.

\bibitem[{Wheeler and Kiladis(1999)}]{Wheeler:1999}
Wheeler, M. and Kiladis, G.~N.: Convectively coupled equatorial waves: Analysis
  of clouds and temperature in the wavenumber-frequency domain, J. Atmos.
  Sci., 56, 374--399, 1999.

\bibitem[{Wheeler and Weickmann(2001)}]{Wheeler_Weickmann:2001}
Wheeler, M. and Weickmann, K.~M.: Real-time monitoring and prediction of modes
  of coherent synoptic to intraseasonal tropical variability, Mon. Weather Rev,
  129, 2677--2694, 2001.

\end{thebibliography}

\end{document}